\documentclass[11pt]{article}
\usepackage{times}
\usepackage{geometry}
\usepackage[utf8]{inputenc}
\usepackage{enumitem,amssymb}
\usepackage{ragged2e}
\usepackage{wrapfig}
\usepackage{graphicx}
\usepackage{xcolor}
\usepackage{orcidlink}
\newlist{thematic}{itemize}{8}
\setlist[thematic]{label=$\square$}
\usepackage{pifont}

\begin{document}
\sloppy 
%\raggedright
{\LARGE ESO Expanding Horizon White Paper: Revealing the properties of matter at supranuclear densities with  gravitational waves}\\ 

{Tim Dietrich$^{1,2}$ \orcidlink{0000-0003-2374-307X}},
{Tanja Hinderer$^{3}$ \orcidlink{0000-0002-3394-6105}},
{Micaela Oertel$^{4,5}$\orcidlink{0000-0002-1884-8654}}, 
{Conrado Albertus Torres$^{6}$\orcidlink{0000-0002-0248-8260}},
{Nils Andersson$^{7}$\orcidlink{0000-0001-8550-3843}},
{Dániel Barta$^{8}$\orcidlink{0000-0001-6841-550X}},
{Andreas Bauswein$^{9}$\orcidlink{0000-0001-6798-3572}},
{Béatrice Bonga$^{10,11}$\orcidlink{0000-0002-5808-9517}},
{Marica Branchesi$^{12}$\orcidlink{0000-0003-1643-0526}},
{G.\ Fiorella Burgio$^{13}$\orcidlink{0000-0003-2195-5693}},
{Stefano Burrello$^{14}$\orcidlink{0000-0002-1132-4073}},
{Prasanta Char$^{6}$\orcidlink{0000-0001-6592-6590}},
{Sylvain Chaty$^{15}$\orcidlink{0000-0002-5769-8601}},
{Maria Colonna$^{14}$\orcidlink{0000-0003-1133-0905}},
{Daniela Doneva$^{16}$\orcidlink{0000-0001-6519-000X}},
{Anthea F. Fantina$^{17}$\orcidlink{0000-0003-2225-4100}},
{Tobias Fischer$^{18,19}$\orcidlink{0000-0003-2479-344X}},
{Juan Garcia-Bellido$^{20}$\orcidlink{0000-0002-9370-8360}},
{Archisman Ghosh$^{21}$\orcidlink{0000-0003-0423-3533}},
{Bruno Giacomazzo$^{22,23}$\orcidlink{0000-0002-6947-4023}},
{Fabian Gittins$^{3}$\orcidlink{0000-0002-9439-7701}},
{Vanessa Graber$^{24}$\orcidlink{0000-0002-6558-1681}},
{Francesca Gulminelli$^{25}$\orcidlink{0000-0003-4354-2849}},
{Jan Harms$^{12,26}$},
{Kostas Kokkotas$^{27}$},
{Felipe J. Llanes-Estrada$^{28}$\orcidlink{0000-0002-2565-4516}},
{Michele Maggiore$^{29}$},
{Gabriel Martinez-Pinedo$^{9}$\orcidlink{0000-0002-3825-0131}},
{Andrea Maselli$^{12}$\orcidlink{0000-0001-8515-8525}},
{Chiranjib Mondal$^{30}$\orcidlink{0000-0002-9238-6144}},
{Samaya Nissanke$^{1,31}$\orcidlink{0000-0001-6573-7773}},
{M Ángeles Pérez García$^{6}$\orcidlink{0000-0003-3355-3704}},
{Cristiano Palomba$^{32}$\orcidlink{0000-0002-4450-9883}},
{Pantelis Pnigouras$^{33}$\orcidlink{0000-0003-1895-9431}},
{Anna Puecher$^{1}$\orcidlink{0000-0003-1357-4348}},
{Michele Punturo$^{34}$},
{Adriana R. Raduta$^{35}$\orcidlink{0000-0001-8421-2040}},
{Violetta Sagun$^{7}$\orcidlink{0000-0001-5854-1617}},
{Armen Sedrakian$^{36,37}$\orcidlink{0000-0001-9626-2643}},
{Nikolaos Stergioulas$^{38}$\orcidlink{0000-0002-5490-5302}},
{Laura Tolos$^{39,40}$\orcidlink{0000-0003-2304-7496}},
{Kadri Yakut$^{41}$\orcidlink{0000-0003-2380-9008}},
{Stoytcho Yazadjiev$^{42}$\orcidlink{0000-0002-1280-9013}}

\begin{footnotesize}
$^{1}$ Institut f\"ur Physik und Astronomie, Universit\"at Potsdam, Haus 28, Karl-Liebknecht-Str. 24/25, 14476, Potsdam, Germany \newline
$^{2}$ Max Planck Institute for Gravitational Physics (Albert Einstein Institute), Am M\"{u}hlenberg 1, Potsdam 14476, Germany \newline
$^{3}$ Institute for Theoretical Physics, Utrecht University, Princetonplein 5, 3584 CC Utrecht, The Netherlands \newline
$^{4}$ Observatoire astronomique de Strasbourg, CNRS, Université de Strasbourg, 11 rue de l'Université, 67000 Strasbourg, France \newline
$^{5}$ Observatoire de Paris, CNRS, Université PSL, 5 place Jules Janssen, 92915 Meudon, France \newline
$^{6}$ Department of Fundamental Physics \& IUFFyM Plaza de la Merced s/n E-37008 Salamanca, Spain \newline
$^{7}$ Mathematical Sciences and STAG Research Centre, University of Southampton, Southampton SO17 1BJ, United Kingdom \newline
$^{8}$ HUN-REN Wigner Research Centre for Physics,
Konkoly-Thege Miklos ut 29-33, 1121 Budapest, Hungary \newline
$^{9}$ GSI Helmholtzzentrum f{\"u}r Schwerionenforschung, Planckstra{\ss}e 1, D-64291 Darmstadt, Germany \newline
$^{10}$ Institute for Mathematics, Astrophysics and Particle Physics, Radboud University, 6525 AJ Nijmegen, The Netherlands \newline
$^{11}$ Theoretical Sciences Visiting Program, Okinawa Inst.\ of Science and Technology Graduate University, Onna, 904-0495, Japan \newline
$^{12}$ Gran Sasso Science Institute, I-67100, L’Aquila (AQ), Italy \newline
$^{13}$ INFN Sezione di Catania, Via S. Sofia 64, 95123 Catania, Italy \newline
$^{14}$ INFN - Laboratori Nazionali del Sud Via S. Sofia, 62, 95123 Catania, Italy \newline
$^{15}$ Université Paris Cité, CNRS, Astroparticule et Cosmologie, F-75013 Paris, France \newline
$^{16}$ Departamento de Astronomia y Astrofisica, Universitat de Valencia, Dr. Moliner 50, 46100, Burjassot (Valencia) Spain \newline
$^{17}$ Grand Accélérateur National d’Ions Lourds, CEA/DRF - CNRS/IN2P3, Boulevard Henri Becquerel, 14076 Caen, France \newline
$^{18}$ Institute of Theoretical Physics, Wroclaw University of Science and Technology, W. Wyspianskiego 27, 50-370 Wroclaw, Poland

$^{19}$ Research Center for Comp.\ Physics and Data Proc., Silesian University, Bezručovo nám.\ 13, CZ-746-01 Opava, Czech Republic \newline
$^{20}$ Department of Theoretical Physics, Universidad Autónoma de Madrid, 28049 Madrid, Spain \newline
$^{21}$ Department of Physics and Astronomy, Ghent University, 
Proeftuinstraat 85, B-9000 Ghent, Belgium \newline
$^{22}$ Dipartimento di Fisica G. Occhialini, Università degli Studi di Milano-Bicocca, Piazza della Scienza 3, I-20126 Milano, Italia \newline
$^{23}$ INFN, Sezione di Milano-Bicocca, Piazza della Scienza 3, I-20126, Milano, Italia \newline
$^{24}$ Department of Physics, Royal Holloway, University of London, Egham Hill Egham, TW20 0EX, United Kingdom \newline
$^{25}$ Université de Caen Normandie, ENSICAEN, CNRS/IN2P3, LPC Caen UMR6534, F-14000 Caen, France \newline
$^{26}$ INFN, Laboratori Nazionali del Gran Sasso, 67100 Assergi, Italy \newline
$^{27}$ Theoretical Astrophysics, University of Tuebingen, Tuebingen 72076, Germany \newline
$^{28}$ Depto. Física Teórica \& IPARCOS, Univ. Complutense de Madrid, Plaza de las Ciencias 1, 28040 Madrid Spain \newline
$^{29}$ Département de Physique Théorique, Université de Genève, 24 quai Ernest Ansermet, 1211 Genève 4, Switzerland and Gravitational Wave Science Center (GWSC), Université de Genève, CH-1211 Geneva, Switzerland \newline
$^{30}$ Universite Libre de Bruxelles, Avenue F. Roosevelt 50, CP 226, 1050 Bruxelles, Belgium \newline
$^{31}$ DESY and the German Centre for Astrophysics (DZA), Platanenallee 6, 15738 Zeuthen, Germany \newline
$^{32}$ INFN, Sezione di Roma I-00185 Roma, Italy \newline
$^{33}$ Department of Physics, University of Alicante, 03690 San Vicente del Raspeig (Alicante), Spain \newline
$^{34}$ Istituto Nazionale di Fisica Nucleare, sezione di Perugia, via Pascoli, 06123 Perugia, Italy \newline
$^{35}$ National Institute for Physics and Nuclear Engineering (IFIN-HH), RO-077125, Bucharest, Romania \newline
$^{36}$ Frankfurt Institute for Advanced Studies, 60438 Frankfurt am Main, Germany \newline
$^{37}$ University of Wroclaw, Wroclaw 50204, Poland \newline
$^{38}$ Department of Physics, Aristotle University of Thessaloniki, 54124 Thessaloniki, Greece \newline
$^{39}$ Institute of Space Sciences (ICE, CSIC), Campus UAB, Carrer de Can Magrans, 08193 Barcelona, Spain \newline
$^{40}$ Institut d’Estudis Espacials de Catalunya (IEEC), 08860 Castelldefels (Barcelona), Spain \newline
$^{41}$ Department of Astronomy and Space Sciences, Faculty of Science,
Ege University, 35100, Izmir, Turkey \newline
$^{42}$ Department of Theoretical Physics, Sofia University `St. Kliment Ohridski' 5 J. Bourchier Blvd. Sofia 1164, Bulgaria 
\newline
\end{footnotesize}
\newpage 

\textbf{Understanding dense matter under extreme conditions is one of the most fundamental puzzles in modern physics. Complex interactions give rise to emergent, collective phenomena. While nuclear experiments and Earth-based colliders provide valuable insights, much of the quantum chromodynamics (QCD) phase diagram at high density and low temperature remains accessible only through astrophysical observations of neutron stars, neutron star mergers, and stellar collapse. Thus, astronomical observations offer a direct window to the physics on subatomic scales with gravitational waves presenting an especially clean channel. \\
Next-generation gravitational-wave observatories, such as the Einstein Telescope, would serve as unparalleled instruments to transform our understanding of neutron star matter. They will enable the detection of up to tens of thousands of binary neutron-star and neutron-star–black-hole mergers per year, a dramatic increase over the few events accessible with current detectors. They will provide an unprecedented precision in probing cold, dense matter during the binary inspiral, exceeding by at least an order of magnitude what current facilities can achieve. Moreover, these observatories will allow us to explore uncharted regimes of dense matter at finite temperatures produced in a subset of neutron star mergers, areas that remain entirely inaccessible to current instruments. Together with multimessenger observations, these measurements will significantly deepen our knowledge of dense nuclear matter.}
\linebreak

\begin{wrapfigure}[22]{r}{0.62\textwidth}
\vspace*{-1.2cm}
\begin{center}
\includegraphics[width=0.61\textwidth]{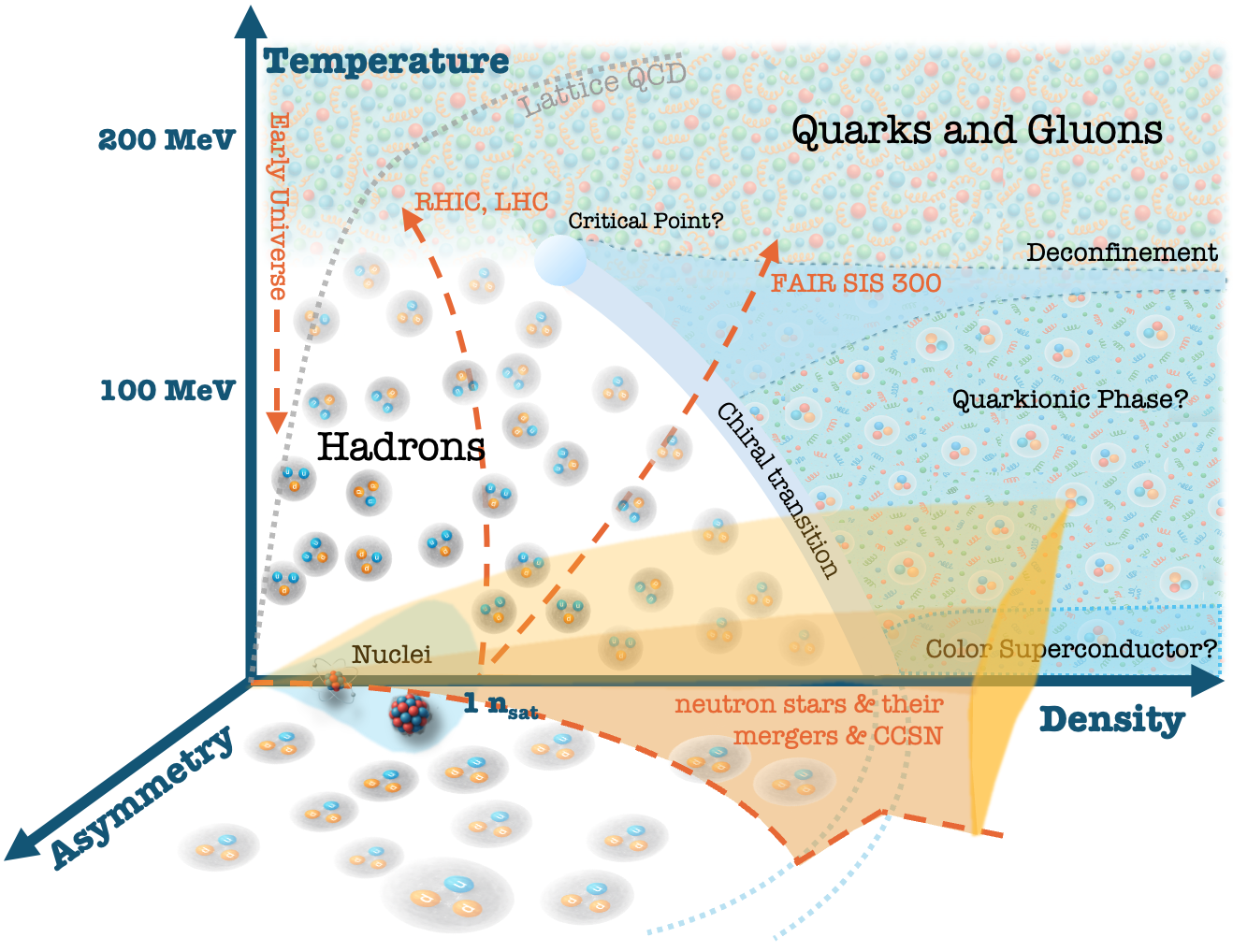}
\caption{{\it Phase diagram of QCD}. Orange shading indicates the range of thermodynamic conditions spanned by NSs, NS mergers, core-collapse supernovae (CCSN), and proto-NSs, all of which are accessible with GWs. Other parts of the diagram are explored in various terrestrial facilities.
}
\label{fig:phasediagram}
\end{center}
\end{wrapfigure}

Despite significant progress in understanding matter under extreme conditions—such as high densities, temperatures, isospin asymmetries (i.e., the neutron to proton ratio) or magnetic fields—many fundamental questions remain unresolved and many regions of the QCD phase diagram (Fig.~\ref{fig:phasediagram}) are  unexplored.  
While major Earth-based collider facilities have substantially advanced our knowledge about strongly-interacting matter in recent years, they probe only limited regimes, e.g., they can not probe low temperature regions and are unable to reach baryon densities beyond a few times nuclear saturation density. In contrast, the extreme environments found in neutron stars (NSs) and core-collapse supernovae offer unique access to such conditions. Observing the gravitational waves (GWs) from the dynamics of such extraordinary objects enables us to probe matter at supranuclear densities. In particular, it enables addressing long-standing questions about the composition of NS interiors: do novel states of hyperonic matter, mesonic condensates, or even (phase) transitions to deconfined quark matter appear?

\begin{wrapfigure}[25]{r}{0.6\textwidth}
\vspace*{-0.8cm}
\begin{center}
\includegraphics[width=0.55\textwidth]{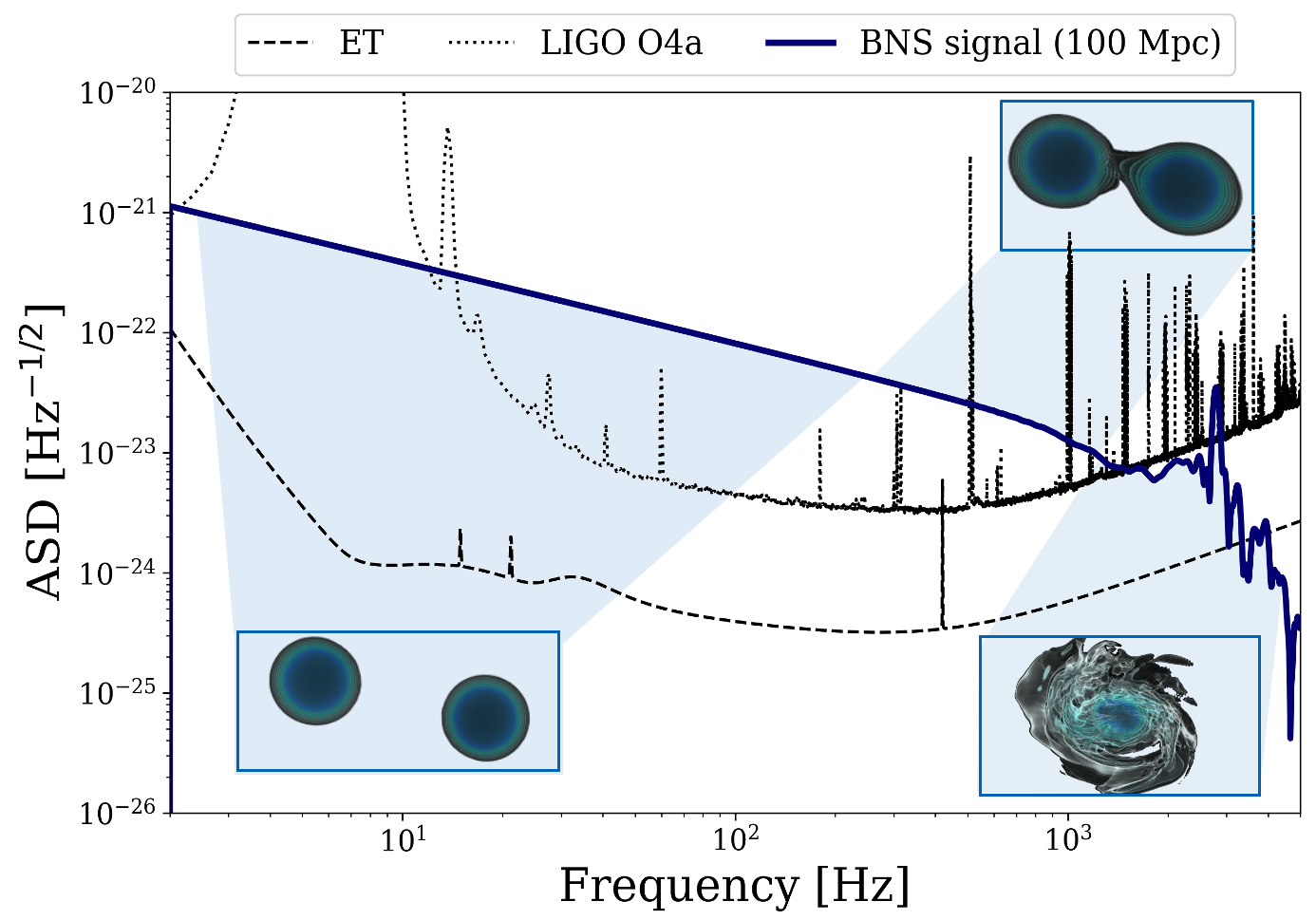}
\caption{{\it Binary NS coalescence seen by GW detectors}. The blue curve is an averaged GW strain amplitude, and snapshots of the density from numerical-relativity simulations indicate interesting regimes: (i) early inspiral, where resonant tidal excitations of NS oscillation modes with low frequency could reveal detailed information about NS interiors, (ii) tidal deformations that encode information on the EOS, and (iii) the postmerger phase in which the finite-temperature equation of state can be probed. Black curves indicate the most recent sensitivity of LIGO (dotted) and of ET (dashed).}
\label{fig:BNS}
\end{center}
\end{wrapfigure}

The astrophysics community has already made significant strides in constraining the equation of state (EOS) of NS matter, however, current constraints remain too imprecise to answer fundamental questions. Recent highlights include: (i) radio pulsar mass measurements providing lower bounds on the EOS-dependent maximum NS mass, e.g., \cite{Fonseca:2021wxt}; (ii) mass and radius determinations from X-ray pulse profile observations, e.g., \cite{Miller:2019cac,Riley:2019yda}
and (iii) the first GW multi-messenger detection of a binary NS merger, GW170817~\cite{LIGOScientific:2017vwq}. The latter has helped establish GWs as a powerful tool for exploring dense matter because of their ability to travel unimpeded from extreme astrophysical environments. It also demonstrated the rich scientific potential of GW astronomy, particularly when combined with electromagnetic observations~\cite{LIGOScientific:2017ync}. The detection of a kilonova and gamma-ray burst associated with the GW event confirmed NS collisions as a site for heavy element (r-process) production and provided constraints on NS tidal deformabilities and thereby the EOS. However, current detectors have limited sensitivity, and mainly probe the inspiral phase of binary NS mergers, where matter remains cold and near equilibrium; cf.~Fig.~\ref{fig:BNS}. 
Planned upgrades to the LIGO, Virgo, and KAGRA detector network are expected to increase the number of gravitational-wave detections and improve measurement accuracy, however, the network will still be limited to relatively few detections and only moderate precision in key parameters.
%Upcoming improvements in the LIGO, Virgo, and KAGRA networks are expected to deliver more detections and enable better constraints, including a broader sampling of NS populations and more precise tidal deformability measurements. However, only future instruments of the third generation such as the Einstein Telescope (ET)~\cite{Punturo:2010zz,Branchesi:2023mws,ET:2025xjr} will reach sensitivities which allow to detect the large number of NS binary systems expected at redshifts corresponding to the peak of star formation. 

Crucial aspects of NS physics will remain inaccessible with current facilities. In particular, the post-merger phase of binary NS collisions remains an uncharted frontier. Depending on properties of the merging NSs, mainly their masses, the collision may result in prompt collapse to a black hole or a hypermassive or supramassive NS that survives for milliseconds to seconds. In the latter case, the remnants feature extreme temperatures, fast (differential) rotation, and violent oscillations, all occurring far from chemical equilibrium. Such physical conditions can neither be reliably described by ab initio QCD calculations nor low-energy nuclear models. However, GW signals from this phase encode invaluable information about the EOS under these conditions, its thermal properties, out-of-equilibrium effects, and possible phase transitions. Current detectors lack the sensitivity required to observe these high-frequency signals, underscoring the need for next-generation observatories such as the Einstein Telescope (ET)~\cite{Punturo:2010zz,Branchesi:2023mws,ET:2025xjr}.

Core-collapse supernovae and isolated NSs offer additional opportunities to probe dense matter under yet different conditions~\cite{ET:2025xjr}. A core-collapse supernova occurring within the Milky Way would allow simultaneous detection of GWs, an electromagnetic signal and neutrinos, providing insight into the explosion mechanism and the properties of hot, dense matter in the proto-NS immediately following collapse. Likewise, continuous GWs from surface deformations on rotating NSs could reveal the properties of NS crusts, magnetic field structures, and elastic stresses, offering an important complement to binary NS observations.

Only next-generation GW observatories such as the ET are poised to transform our ability to probe extreme matter. They will offer an order-of-magnitude improvement in sensitivity across a wide frequency range. This improvement will make it possible to observe up to tens of thousands of NS mergers per year, of which a small fraction will have signal-to-noise ratios of more than several hundred, which will allow precise measurements of NS masses and tidal deformability. Subdominant effects such as excitations of characteristic modes during inspiral, enabling astereoseismology of NS interiors and hence potentially directly probing strong phase transitions or composition gradients, might also be measurable by ET for a small subset of events. Moreover, for some events, GWs from tidal disruptions of NSs with black hole companions will be observable, yielding additional EOS information. Crucially, third-generation observatories will, for the first time, have the capabilities to detect post-merger GW signals from close binary NS collisions, which for a few of the events will be sufficiently loud to provide access to the hot, non-equilibrium phases of NS matter. In addition to binary mergers, ET will significantly enhance the prospects of observing GWs from isolated NSs and galactic supernovae, with the latter offering another possibility for unprecedented multi-messenger science.  

In general, the increased sensitivity of ET will enable rich multi-messenger observations, with joint GW, electromagnetic (with promising projects such as the eXTP~\cite{Li:2025uaw} and SKA \cite{SKAPulsarScienceWorkingGroup:2025ins}), and neutrino data to deliver a comprehensive view of dense matter across a broad range of physical conditions. Observations by ET and other third-generation GW detectors will unveil details about heavy element synthesis, constrain the EOS across a range of temperatures and compositions, and potentially use NS mergers as probes for exotic physics beyond the Standard Model. With strong support from nuclear theory, ET observations could further be used to test General Relativity and the coupling of gravity to matter in strong-field conditions and investigate potential dark matter effects in compact stars~\cite{ET:2025xjr}. Additional unique multi-messenger opportunities with ET for profound insights on dark energy, gravity, and chemical enrichment of the universe are discussed in a companion White Paper about multi-messenger science.

In summary, ET represents a transformative opportunity for nuclear and astrophysics. With its unprecedented sensitivity, ET will enable precision measurements of the EOS under both cold and hot conditions, map NS merger dynamics, and provide insight into the interior structure of NSs. By combining GW data with multimessenger observations and laboratory experiments, ET promises to elucidate longstanding questions about dense matter, phase transitions, and the behavior of the strong force. Through coordinated theoretical, observational, and experimental efforts, ET can provide an almost complete view on matter under extreme conditions and open new frontiers at the intersection of astrophysics, nuclear physics, and fundamental physics.\\

\begin{footnotesize}
\bibliographystyle{plain}
\bibliography{refs.bib}

@article{LIGOScientific:2017vwq,
    author = "Abbott, B. P. and others",
    collaboration = "LIGO Scientific, Virgo",
    title = "{GW170817: Observation of Gravitational Waves from a Binary Neutron Star Inspiral}",
    eprint = "1710.05832",
    archivePrefix = "arXiv",
    primaryClass = "gr-qc",
    reportNumber = "LIGO-P170817",
    doi = "10.1103/PhysRevLett.119.161101",
    journal = "Phys. Rev. Lett.",
    volume = "119",
    number = "16",
    pages = "161101",
    year = "2017"
}

@article{SKAPulsarScienceWorkingGroup:2025ins,
    author = "Basu, Avishek and others",
    collaboration = "SKA Pulsar Science Working Group",
    title = "{Probing neutron star interiors and the properties of cold ultra-dense matter with the SKAO}",
    eprint = "2512.16162",
    archivePrefix = "arXiv",
    primaryClass = "astro-ph.HE",
    month = "12",
    year = "2025"
}

@article{Branchesi:2023mws,
    author = "Branchesi, Marica and others",
    title = "{Science with the Einstein Telescope: a comparison of different designs}",
    eprint = "2303.15923",
    archivePrefix = "arXiv",
    primaryClass = "gr-qc",
    reportNumber = "ET-0084A-23",
    doi = "10.1088/1475-7516/2023/07/068",
    journal = "JCAP",
    volume = "07",
    pages = "068",
    year = "2023"
}

@article{Punturo:2010zz,
    author = "Punturo, M. and others",
    editor = "Ricci, Fulvio",
    title = "{The Einstein Telescope: A third-generation gravitational wave observatory}",
    doi = "10.1088/0264-9381/27/19/194002",
    journal = "Class. Quant. Grav.",
    volume = "27",
    pages = "194002",
    year = "2010"
}

@article{LIGOScientific:2017ync,
    author = "Abbott, B. P. and others",
    collaboration = "LIGO Scientific, Virgo, Fermi GBM, INTEGRAL, IceCube, AstroSat Cadmium Zinc Telluride Imager Team, IPN, Insight-Hxmt, ANTARES, Swift, AGILE Team, 1M2H Team, Dark Energy Camera GW-EM, DES, DLT40, GRAWITA, Fermi-LAT, ATCA, ASKAP, Las Cumbres Observatory Group, OzGrav, DWF (Deeper Wider Faster Program), AST3, CAASTRO, VINROUGE, MASTER, J-GEM, GROWTH, JAGWAR, CaltechNRAO, TTU-NRAO, NuSTAR, Pan-STARRS, MAXI Team, TZAC Consortium, KU, Nordic Optical Telescope, ePESSTO, GROND, Texas Tech University, SALT Group, TOROS, BOOTES, MWA, CALET, IKI-GW Follow-up, H.E.S.S., LOFAR, LWA, HAWC, Pierre Auger, ALMA, Euro VLBI Team, Pi of Sky, Chandra Team at McGill University, DFN, ATLAS Telescopes, High Time Resolution Universe Survey, RIMAS, RATIR, SKA South Africa/MeerKAT",
    title = "{Multi-messenger Observations of a Binary Neutron Star Merger}",
    eprint = "1710.05833",
    archivePrefix = "arXiv",
    primaryClass = "astro-ph.HE",
    reportNumber = "LIGO-P1700294, VIR-0802A-17, FERMILAB-PUB-17-478-A-AE-CD",
    doi = "10.3847/2041-8213/aa91c9",
    journal = "Astrophys. J. Lett.",
    volume = "848",
    number = "2",
    pages = "L12",
    year = "2017"
}

@article{ET:2025xjr,
    author = "Abac, Adrian and others",
    collaboration = "ET",
    title = "{The Science of the Einstein Telescope}",
    journal = "arXiv:2503.12263",
    archivePrefix = "arXiv",
    primaryClass = "gr-qc",
    reportNumber = "ET-0036C-25",
    month = "3",
    year = "2025"
}

@article{Fonseca:2021wxt,
    author = "Fonseca, E. and others",
    title = "{Refined Mass and Geometric Measurements of the High-mass PSR J0740+6620}",
    eprint = "2104.00880",
    archivePrefix = "arXiv",
    primaryClass = "astro-ph.HE",
    doi = "10.3847/2041-8213/ac03b8",
    journal = "Astrophys. J. Lett.",
    volume = "915",
    number = "1",
    pages = "L12",
    year = "2021"
}

@article{Miller:2019cac,
    author = "Miller, M. C. and others",
    title = "{PSR J0030+0451 Mass and Radius from $NICER$ Data and Implications for the Properties of Neutron Star Matter}",
    eprint = "1912.05705",
    archivePrefix = "arXiv",
    primaryClass = "astro-ph.HE",
    doi = "10.3847/2041-8213/ab50c5",
    journal = "Astrophys. J. Lett.",
    volume = "887",
    number = "1",
    pages = "L24",
    year = "2019"
}

@article{Riley:2019yda,
    author = "Riley, Thomas E. and others",
    title = "{A $NICER$ View of PSR J0030+0451: Millisecond Pulsar Parameter Estimation}",
    eprint = "1912.05702",
    archivePrefix = "arXiv",
    primaryClass = "astro-ph.HE",
    doi = "10.3847/2041-8213/ab481c",
    journal = "Astrophys. J. Lett.",
    volume = "887",
    number = "1",
    pages = "L21",
    year = "2019"
}

@article{Li:2025uaw,
    author = "Li, Ang and others",
    title = "{Dense matter in neutron stars with eXTP}",
    eprint = "2506.08104",
    archivePrefix = "arXiv",
    primaryClass = "astro-ph.HE",
    doi = "10.1007/s11433-025-2761-4",
    journal = "Sci. China Phys. Mech. Astron.",
    volume = "68",
    number = "11",
    pages = "119503",
    year = "2025"
}
\end{footnotesize}
\end{document}